\documentclass[aps,prb,groupedaddress,twocolumn]{revtex4}
\usepackage{amssymb}
\usepackage{dcolumn}

\usepackage{graphicx}

\def\be{\begin{equation}}
\def\ee{\end{equation}}
\def\ba{\begin{eqnarray}}
\def\ea{\end{eqnarray}}
\begin{document}

\title{Two ultracold atoms in a completely anisotropic trap}
\author{Jun-Jun Liang$^{1,2}\thanks{%
Corresponding author$^{\prime }$s E-mail address: liangjj@sxu.edu.cn}$,Chao
Zhang$^1$}
\address{$^1$Department of Physics, Shanxi University, Taiyuan 030006, China\\
$^2$Institute of Theoretical Physics, Shanxi University, Taiyuan
030006, China}

\begin{abstract}
As a limiting case of ultracold atoms trapped in deep optical
lattices, we consider two interacting atoms trapped in a general
anisotropic harmonic oscillator potential, and obtain exact
solutions of the Schr\"{o}dinger equation for this system. The
energy spectra for different geometries of the trapping potential
are compared.

\end{abstract}

\maketitle

 PACS number(s): 34.50. s, 32.80.Pj

Recently optical lattice becomes a convenient tool to study many-body
physics in periodical potential, and the physics of quantum degenerate
atomic gases trapped in an optical lattice have been intensively
investigated theoretically\cite{Dickerscheid,Koetsier,Gubbels,Diener} and
experimentally\cite{Sto¡§ferle,Deuretzbacher,C. Ospelkaus}. The
characteristic of short-range interaction between atoms makes optical
lattice ideal to experimentally realize Bose (or Fermi) Hubbard models. In
deep optical lattice multiband Hubbard models are extremely difficult to
handle. One can gain some insight into this system by neglecting tunneling
between adjacent sites. In this case, Bloch bands are nearly flat, and an
individual well may be approximated by a harmonic oscillator potential. As a
result, we model the lattice as an array of microscopic harmonic traps, each
of which is occupied with a few atoms. A precise understanding of atoms
interacting in this simple model is a prerequisite for analysis of many-body
physics in optical lattices with resonantly enhanced interactions. We only
consider the physics of two identical atoms now, instead of different atoms
that may feel different trapping frequencies, which leads to a coupling of
center-of-mass (CM) and relative motion \cite{Deuretzbacher}.

The system of two interacting atoms in a harmonic trap has been studied
analytically both in spherically \cite{Deuretzbacher,FoP1998} and axially
symmetric cases \cite{Idziaszek}. The interaction is described in terms of
an s-wave pseudopotential by the regularized $\delta $- function\cite{Huang}%
. Higher partial wave, such as p-wave is also considered\cite{Stock}. In
this paper we investigate the completely anisotropic harmonic potential and
obtain the exact solutions for two interacting atoms confined in a trap with
three different frequencies. At last, we compare the properties of energy
spectra for trapping potential of different geometries.

We consider two interacting atoms with identical mass $m$ , which are
confined in a completely anisotropic harmonic trapping potential with three
frequencies $\omega _x$, $\omega _y$ and $\omega _z$. The Hamiltonian can be
written as

\begin{equation}
H=-\frac{\hbar ^2}{2m}\nabla _1^2-\frac{\hbar ^2}{2m}\nabla _2^2+V_{Trap}(%
{\bf r}_1)+V_{Trap}({\bf r}_2)+V_{int}({\bf r}_1-{\bf r}_2)\text{,}
\label{Hamiltonian}
\end{equation}
where ${\bf r}_1$ and ${\bf r}_2$ denote the positions of two atoms
respectively, and $V_{Trap}({\bf r})$ is the completely anisotropic harmonic
trapping potential

\begin{equation}
V_{Trap}\left( {\bf r}\right) =\frac 12m(\omega _x^2x^2+\omega
_y^2y^2+\omega _z^2z^2)\text{.}
\end{equation}

In the ultracold regime, atomic interaction is dominated by $s$-wave
scattering. In this case, the interaction potential is modeled by a
pointlike form which is expressed in terms of the so-called regularized $%
\delta $-function
\begin{equation}
V_{int}({\bf r})=\frac{4\pi a_0\hbar ^2}m\delta \left( {\bf r}\right) \frac %
\partial {\partial r}r\text{,}
\end{equation}
where $r=\left| {\bf r}\right| $ denotes the distance between two atoms and $%
a_0$ the scattering length.

For convenience, in the following calculations we use dimensionless
variables, in which lengths are expressed in units of $\sqrt{\hbar /m\omega
_x}$ and energies are expressed in units of $\hbar \omega _x$. In addition,
all frequencies are denoted with $\omega _x$ by introducing parameters $\eta
_y$ and $\eta _z$, where $\eta _y=\omega _y/\omega _x$ and $\eta _z=\omega
_z/\omega _x$. In this two-body system, as the Hamiltonian (\ref{Hamiltonian}%
) has a quadratic form for both kinetic energy and harmonic oscillator
potential, the motion of CM and relative one can be decoupled by introducing
${\bf r}=({\bf r}_1-{\bf r}_2)\strut /\sqrt{2}$ and ${\bf R}=({\bf r}_1+{\bf %
r}_2)/\sqrt{2}$, where ${\bf R}$ is the CM coordinate and ${\bf r}$ the
relative coordinate. The eigenfunctions and eigenvalues of CM Hamiltonian $%
H_{CM}=-\frac 12\nabla _R^2+V_{CM}({\bf R})$, where $V_{CM}({\bf R}%
)=(X^2+\eta _y^2Y^2+\eta _z^2Z^2)/2$, are the solutions of three-dimension
harmonic oscillator, which are tackled analytically in the standard textbook
of quantum mechanics.

Now we consider the relative motion of the system, of which the Hamiltonian $%
H_{rel}$ reads

\begin{equation}
H_{rel}=-\frac 12\nabla ^2+V_{rel}({\bf r})+\sqrt{2}\pi a_0\delta ({\bf r})%
\frac \partial {\partial r}r\text{,}  \label{relative}
\end{equation}
where $V_{rel}({\bf r})=(x^2+\eta _y^2y^2+\eta _z^2z^2)/2$. The solutions
can be determined by the following Schr\"{o}dinger equation

\begin{equation}
H_{rel}\Psi =E\Psi \text{.}  \label{shrodinger}
\end{equation}

To obtain the eigenvalues and eigenfunctions of the Hamiltonian (\ref
{relative}) is our main tasks in this paper. First let us consider two
non-interacting atoms in the same trapping potential, and the system
satisfies the following equation
\begin{eqnarray}
\left[ -\frac 12\nabla ^2+\frac 12(x^2+\eta _y^2y^2+\eta
_z^2z^2)\right] \Phi
_{n_xn_yn_z}(x,y,z)\nonumber\\=E_{n_xn_yn_z}\Phi
_{n_xn_yn_z}(x,y,z)\text{,} \label{noninteract}
\end{eqnarray}
with $\Phi _{n_xn_yn_z}(x,y,z)=\Phi _{n_x}(x)\Phi _{n_y}(\sqrt{\eta _y}%
y)\Phi _{n_z}(\sqrt{\eta _z}z)$, in which $\Phi _{n_w}(\alpha w)=N_{n_w}e^{-%
\frac 12\alpha ^2w^2}H_{n_w}(\alpha w)$ denotes the eigenfunctions of 1D
harmonic oscillator, where the constant $N_{n_w}=(\alpha /\sqrt{\pi }%
2^{n_w}(n_w)!)^{1/2}$ with parameter $\alpha =1,$ $\sqrt{\eta _y},\sqrt{\eta
_z}$ and corresponding variable $w=x,y,z$, while $H_{n_w}(\alpha w)$ is the
Hermite function. The dimensionless eigenenergies of the system are $%
E_{n_xn_yn_z}=$ $(n_x+1/2)+\eta _y(n_y+1/2)+\eta _z(n_z+1/2)$. We use
complete set $\left\{ \Phi _{n_xn_yn_z}(x,y,z)\right\} $ to expand the
unknown wavefunction $\Psi $ in Eq.(\ref{shrodinger}) ,

\begin{equation}
\Psi ({\bf r})=\sum\limits_{n_x\text{,}n_y\text{,}n_z}c_{n_xn_yn_z}\Phi
_{n_xn_yn_z}\text{.}  \label{expansion}
\end{equation}

By inserting (\ref{relative}) and (\ref{expansion}) into (\ref{shrodinger}),
the Schr\"{o}dinger equation can be express as

\begin{eqnarray}
\sum\limits_{n_x\text{,}n_y\text{,}n_z}c_{n_xn_yn_z}(E_{n_xn_yn_z}-E)\Phi
_{n_xn_yn_z}+\sqrt{2}\pi a_0\delta (r)\nonumber\\\times\frac \partial {\partial r}%
r\sum\limits_{n_x\text{,}n_y\text{,}n_z}c_{n_xn_yn_z}\Phi _{n_xn_yn_z}=0%
\text{.}  \label{exp2}
\end{eqnarray}

To determine the expansion coefficients $c_{n_xn_yn_z}$ we project Eq.(\ref
{exp2}) onto state $\Phi _{m_xm_ym_z}$ with nonnegative integers $m_x$, $m_y$
and $m_z$, and obtain the structure of $c_{m_xm_ym_z}$

\begin{equation}
c_{m_xm_ym_z}=C\frac{[\Phi _{m_x}\left( 0\right) \Phi _{m_y}\left( 0\right)
\Phi _{m_z}\left( 0\right) ]^{*}}{E_{m_xm_ym_z}-E}\text{,}  \label{exp3}
\end{equation}
where $C$ is a constant that is related to

\begin{equation}
C=-\sqrt{2}\pi a_0\left[ \frac \partial {\partial r}\left( r\sum\limits_{n_x%
\text{,}n_y\text{,}n_z}c_{n_xn_yn_z}\Phi _{n_xn_yn_z}(x,y,z)\right) \right]
_{r\rightarrow 0}\text{.}  \label{exp4}
\end{equation}

Substituting the expression (\ref{exp3}) for coefficient $c_{n_xn_yn_z}$
into Eq.(\ref{exp4}), the constant C is removed and we have an equation
which denotes the relation between the scattering length and eigenenergies

\begin{equation}
\sqrt{2}\pi \left[ \frac \partial {\partial r}\left( r\Psi _E(x,y,z)\right)
\right] _{r\rightarrow 0}=-\frac 1{a_0}\text{,}  \label{exp5}
\end{equation}
where
\begin{eqnarray}
&&\Psi
_E(x,y,z)=\nonumber\\&&\sum\limits_{n_x\text{,}n_y\text{,}n_z}\frac{\left[
\Phi _{n_x}(0)\Phi _{n_y}(0)\Phi _{n_z}(0)\right] ^{*}\Phi
_{n_x}\left( x\right) \Phi _{n_y}(y)\Phi _{n_z}\left( z\right)
}{E_{n_xn_yn_z}-E} \label{wavefunction0}
\end{eqnarray}
is the non-normalized eigenstates of the Schr\"{o}dinger equation
(\ref {shrodinger}). Using the expressions of eigenfunctions and
eigenvalues of Eq.(\ref{noninteract}) we find the form of
wavefunction

\begin{eqnarray}
\Psi _E(x,y,z) =e^{-\frac 12\left( x^2+\eta _yy^2+\eta _zz^2\right)
}\sum\limits_{n_x\text{,}n_y\text{,}n_z}N_{n_x}^2N_{n_y}^2N_{n_z}^2
\nonumber \\
\times \frac{H_{n_x}\left( x\right) H_{n_x}\left( 0\right)
H_{n_y}\left(
\sqrt{\eta _y}y\right) H_{n_y}\left( 0\right) H_{n_z}\left( \sqrt{\eta _z}%
z\right) H_{n_z}\left( 0\right) }{n_x+\eta _yn_y+\eta _zn_z+f(E)}\text{,}
\end{eqnarray}
where we let $f(E)=\frac 12+\frac 12\eta _y+\frac 12\eta _z-E$ for
convenience and the denominator $n_x+\eta _yn_y+\eta _zn_z+f(E)$ denotes the
shift of energy.

By utilizing integral representation
\begin{equation}
\frac 1{n_x+\eta _yn_y+\eta _zn_z+f(E)}=\int\limits_0^1dqq^{n_x+\eta
_yn_y+\eta _zn_z+f(E)-1}\text{,}
\end{equation}
which is valid for $f(E)>0$ since $n_x+\eta_y n_y+\eta_z n_z
\geqslant 0$, the wavefunction is transformed into

\begin{eqnarray}
&&\Psi _E(x,y,z) =\sum\limits_{n_x\text{,}n_y\text{,}n_z}\frac{(\eta
_y\eta _z)^{1/2}}{\pi ^{3/2}}e^{-\frac 12(x^2+\eta _yy^2+\eta
_zz^2)}\nonumber \\ &&\int\limits_0^1dqq^{f(E)-1} \times \left[
q^{n_x}\frac{H_{n_x}(x)H_{n_x}(0)}{2^{n_x}n_x!}\right]\nonumber \\
&&\times\left[ q^{\eta _yn_y}\frac{H_{n_y}(\sqrt{\eta _y}y)H_{n_y}(0)}{2^{n_y}n_y!}%
\right] \left[ q^{\eta _zn_z}\frac{H_{n_z}(\sqrt{\eta _z}z)H_{n_z}(0)}{%
2^{n_z}n_z!}\right] \text{.}  \label{wavefunction1}
\end{eqnarray}

The limitation $f(E)>0$ effectively means that we restrict the following
steps to energies smaller than the non-interacting harmonic oscillator
ground state energy. For higher energies, analytic continuation may be used.
The series given by the product of two Hermite functions with variable $v$
can be simplified via following formula

\begin{eqnarray}
\sum\limits_{n=0}^\infty&& \frac{H_n\left( z\right) H_n\left( z_1\right) }{n!}%
v^n\nonumber \\&&=\frac 1{\sqrt{1-4v^2}}\exp \left( \frac{2v\left(
2v\left( z^2+z_1^2\right) -2zz_1\right) }{4v^2-1}\right) \text{,}
\label{sum}
\end{eqnarray}
which is valid for $|v|<1/2$. In the expression
(\ref{wavefunction1}) $v=q/2$ satisfies the condition when the
integral variable $q$ changes from 0 to 1. The eigenfunction can be
obtained in integral representation with (\ref{sum})

\begin{eqnarray}
&&\Psi _E(x,y,z)=\frac{(\eta _y\eta _z)^{1/2}}{2\pi ^{3/2}}e^{-\frac 12%
(x^2+\eta _yy^2+\eta _zz^2)}\int\limits_0^\infty dte^{-ut}\nonumber \\
&&\times\frac{\exp (\frac{%
x^2e^{-t}}{e^{-t}-1})}{\sqrt{1-e^{-t}}}\frac{\exp (\frac{\eta
_yy^2e^{-\eta _yt}}{e^{-\eta _yt}-1})}{\sqrt{1-e^{-\eta
_yt}}}\frac{\exp (\frac{\eta _zz^2e^{-\eta _zt}}{e^{-\eta
_zt}-1})}{\sqrt{1-e^{-\eta _zt}}}\text{,} \label{wavefunction}
\end{eqnarray}
where we introduce new variables $t$ and $u$, which are defined as $%
q^2=e^{-t},u=\frac{f(E)}2$.

Due to the pointlike force of two atoms, it is necessary to check the
behavior of the wavefunction for small $x$, $y$ and $z$. When $%
x,y,z\rightarrow 0$, the main contribution for (\ref{wavefunction}) is
dominated by small $t$. It is reasonable to perform the approximation $%
e^{-\lambda t}=1-\lambda t$ for $\lambda =1$, $\eta _y$, $\eta _z$
in the wavefunction and neglect the influence of energy in the
leading order. After some straightforward algebra, the wavefunction
for small $t$ can be written as

\begin{eqnarray}
\Psi _E(x,y,z)&\approx& \frac 1{2\pi ^{3/2}}\int\limits_0^\infty
dt\frac{\exp [-(x^2+y^2+z^2)/t]}{t^{3/2}}\nonumber \\&=&\frac 1{2\pi
(x^2+y^2+z^2)^{1/2}}\text{,}\quad x,y,z\rightarrow 0\text{,}
\end{eqnarray}
which gives no contribution to (\ref{exp5}), and the divergent factor $%
t^{-3/2}$ in the integral above can be subtracted from the wavefunction $%
\Psi _E(x,y,z)$. Using the fact that

\begin{eqnarray}
\frac \partial {\partial r}[rg(x,y,z)]=&&g(x,y,z)+x\frac{\partial g(x,y,z)}{%
\partial x}+y\frac{\partial g(x,y,z)}{\partial y}\nonumber \\
&&+z\frac{\partial g(x,y,z)}{\partial z}\text{,}
\end{eqnarray}
with $g(x,y,z)$ denoting an analytic function, it can be verified
easily that

\begin{equation}
\left[ \frac \partial {\partial r}(r\Psi _E(r))\right] _{r\rightarrow
0}=\left[ \Psi _E(r)\right] _{x,y,z\rightarrow 0}\text{,}  \label{derivation}
\end{equation}
Combining (\ref{wavefunction}) and (\ref{derivation}), we obtain

\begin{eqnarray}
&&\left[ \frac \partial {\partial r}r(\Psi _E(x,y,z))\right] _{r\rightarrow 0}=%
\frac{(\eta _y\eta _z)^{1/2}}{2\pi ^{3/2}}\nonumber\\
&&\times \int\limits_0^\infty dt\left[
\frac{e^{-ut}}{\sqrt{1-e^{-t}}\sqrt{1-e^{-\eta _yt}}\sqrt{1-e^{-\eta _zt}}}-%
\frac 1{t^{3/2}}\right] \text{.}  \label{derivation2}
\end{eqnarray}

\begin{figure}[t]
\begin{center}
\includegraphics[width=0.5\textwidth]{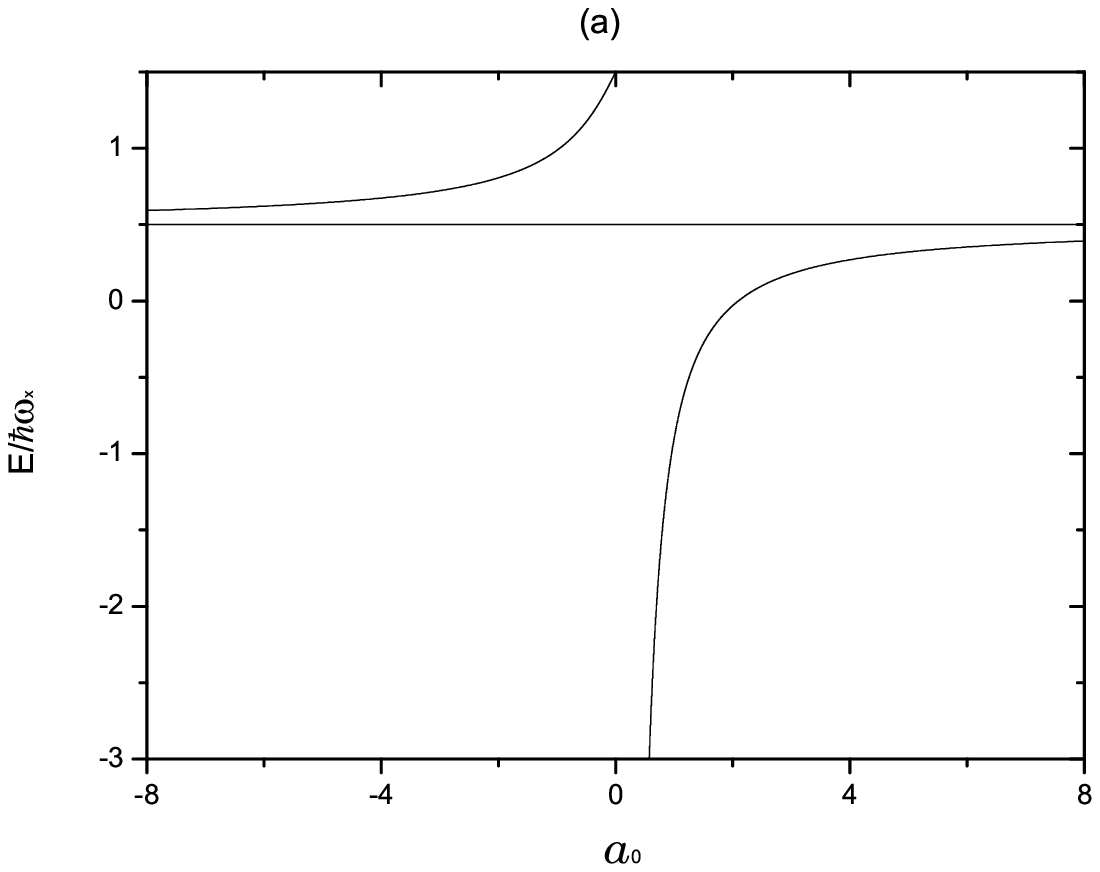}
\includegraphics[width=0.5\textwidth]{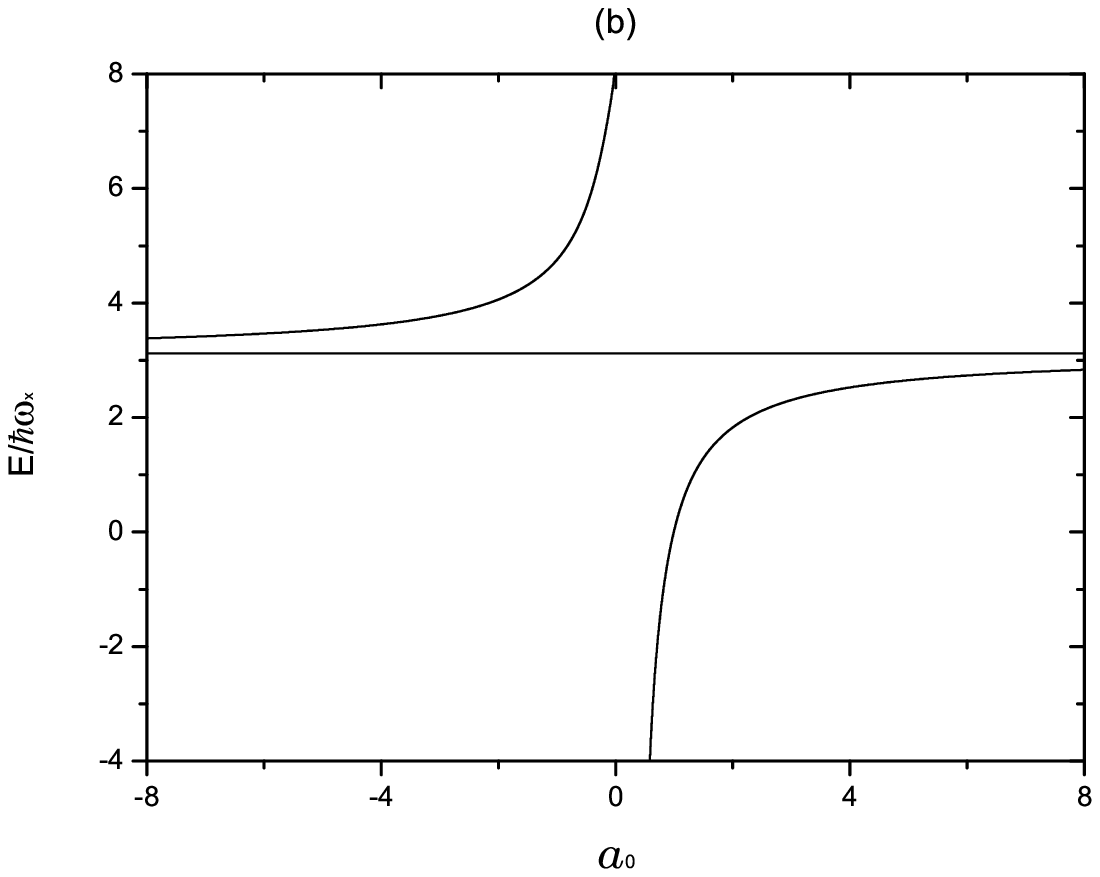}
\end{center}
\caption{Eigenenergies of relative motion for the system of two
atoms interacting via
s-wave pseudopotential and confined in a harmonic trap with parameters (a) $%
\eta _y=1$, $\eta _z=1$ (b) $\eta _y=5$, $\eta _z=10$. The
scattering length $a_0$ is scaled in the units of $\sqrt{\hbar
/m\omega _x}$.} \label{Fig1}
\end{figure}

Substituting the expression (\ref{derivation2}) into (\ref{exp5}), the
eigenenergies takes the form of implicit function

\begin{equation}
Z(u)=-\frac{\sqrt{2\pi }}{a_0}\text{,}  \label{equation4}
\end{equation}
where

\begin{equation}
Z(u)=\int\limits_0^\infty dt\left[ \frac{(\eta _y\eta _z)^{1/2}e^{-ut}}{%
\sqrt{1-e^{-t}}\sqrt{1-e^{-\eta _yt}}\sqrt{1-e^{-\eta _zt}}}-\frac 1{t^{3/2}}%
\right] \text{.}  \label{result1}
\end{equation}

When the trap is axially symmetric, let $\eta _y=\eta _z=\eta $ and (\ref
{equation4}) reduces to

\begin{equation}
\int\limits_0^\infty dt\left[ \frac{\eta e^{-ut}}{\sqrt{1-e^{-t}}(1-e^{-\eta
t})}-\frac 1{t^{3/2}}\right] =-\frac{\sqrt{2\pi }}{a_0}\text{,}
\end{equation}
which is the result given in \cite{Idziaszek}. The additional factor $\sqrt{2%
}$ on the right side is due to different choice of the relative and CM
coordinates.

For the isotropic case, by substituting $\eta _y=\eta _z=1$ into (\ref
{equation4}) the result agrees with the well-known relation between
eigenenergies and scattering length in \cite{FoP1998}

\begin{equation}
\sqrt{2}\frac{\Gamma (u)}{\Gamma (u-\frac 12)}=\frac 1{a_0}\text{.}
\end{equation}

In the case of $f(E)<0$, the eigenenergies are generally determined by
numerical method and we limit our discussion to analytic regime of $f(E)>0$
(i.e. $u>0$).

Fig.1 shows the energy spectra of two interacting ultracold atoms confined
in harmonic traps for (a) isotropic case of $\eta _y=1$, $\eta _z=1$ and (b)
completely anisotropic case of $\eta _y=5$, $\eta _z=10$. When $a_0=0$, the
interaction disappears, and the pole of $Z(u)$ presents the energy of two
atoms, which is corresponding to the ground-state eigenenergy of the
harmonic oscillator with non-interacting case, and this can be verified
easily by (\ref{wavefunction0}), (\ref{derivation}), (\ref{derivation2}) and
(\ref{equation4}). In the isotropic case, the non-interacting ground state
energy is $E=3/2$, and $E=16/2$ in the fully anisotropic case. In the case
of $a_0<0$, the energy level is shifted downwards, while the scattering
exhibits an attractive feature. On the contrary, for $a_0>0$ the energy
approaches $1/a_0^2$ while interactions are repulsive. For the two unitary
limits of $a_0\rightarrow -\infty $ and $a_0\rightarrow +\infty $, they have
the same asymptotic values which become larger with the increase of $\eta _y$
and $\eta _z$, and this is due to the zero of $Z(u)$. It is obvious that
there is no pole of $Z(u)$ for $u>0$. We note that by choosing larger ratios
of frequencies, the energy level moves upwards.

In summary, we have presented the analytical solutions for the system of two
ultracold atoms interacting via $s$-wave pseudopotential in the completely
anisotropic harmonic trap. For special ratios of frequencies, we obtained
the results both in spherically symmetric case\cite{FoP1998} and axially
symmetric one\cite{Idziaszek}. With the model developed in this paper, the
system of two atoms interacting via s-wave pseudopotential can be studied in
any shape of the harmonic trap. Our theoretical result shows a clear
physical picture and can be used in many-body physics of ultracold atoms
trapped in the optical lattice.

This work was supported by the Youth Science Foundations of Shanxi Province
and Shanxi University under Grants Nos. 2006021002, and by the National
Science Foundation of China under Grants Nos. 10444002.

\end{document}